\documentclass[%
 reprint,
superscriptaddress,
showpacs,
 amsmath,amssymb,
 aps,
floatfix,
]{revtex4-1}

\usepackage{dcolumn}
\usepackage{bm}

\usepackage{color}
\usepackage{epsfig}

\usepackage{ulem} 

\begin{document}


\title{Dynamical DMRG study of spin and charge excitations in the four-leg $t$-$t'$-$J$ ladder}

\author{Takami Tohyama}
\email{tohyama@rs.tus.ac.jp}
\affiliation{Department of Applied Physics, Tokyo University of Science, Tokyo 125-8585, Japan}

\author{Michiyasu Mori}
\affiliation{Advanced Science Research Center, Japan Atomic Energy Agency, Tokai, Ibaraki, 319-1195, Japan}

\author{Shigetoshi Sota}
\affiliation{Computational Materials Science Research Team, RIKEN Advanced Institute for Computational Science (AICS), Kobe, Hyogo 650-0047, Japan}


\date{\today}
             

\begin{abstract}

The ground state of the $t$-$t'$-$J$ ladder with four legs favors a striped charge distribution for the parameters corresponding to hole-doped cuprate superconductors. We investigate the dynamical spin and charge structure factors of the model by using the dynamical density matrix renormalization group (DMRG) and clarify the influence of the stripe on the structure factors. The dynamical charge structure factor along the momentum direction from $\mathbf{q}=(0,0)$ to $(\pi,0)$ clearly shows low-energy excitations corresponding to the stripe order in hole doping. On the other hand, the stripe order weakens in electron doping, resulting in fewer low-energy excitations in the charge channel. In the spin channel, we find incommensurate spin excitations near $\mathbf{q}=(\pi,\pi)$ forming an hourglass behavior in hole doping, while in electron doping we find clearly spin-wave-like dispersions starting from  $\mathbf{q}=(\pi,\pi)$. Along the $(0,0)$-$(\pi,0)$ direction, the spin excitations are strongly influenced by the stripes in hole doping, resulting in two branches that form a discontinuous behavior in the dispersion. In contrast, the electron-doped systems show a downward shift in energy toward $(\pi,0)$. These behaviors along the $(0,0)$-$(\pi,0)$ direction are qualitatively similar to momentum-dependent spin excitations recently observed by resonant inelastic x-ray scattering experiments in hole- and electron-doped cuprate superconductors.

\end{abstract}
\maketitle

\section{Introduction}

In cuprate superconductors, spin excitations near the magnetic zone center in the Brillouin zone (BZ) change with hole doping from a spin-wave-type excitation in the Mott insulating phase to an hourglass-type excitation as observed by inelastic neutron scattering (INS)~\cite{Fujita2012}. The formation of charge stripes in hole-doped cuprates~\cite{Tranquada1995} has been assigned to a possible origin of the hourglass-type excitation based on a two-dimensional (2D) single-band Hubbard model with nearest-neighbor hopping $t$, next-nearest-neighbor hopping $t'$, and on-site Coulomb interaction $U$~\cite{Kaneshita2001,Seibold2005,Seibold2006} and on a localized spin model~\cite{Kruger2003,Carlson2004}. Recent quantum Monte Carlo (QMC) calculations of the dynamical spin structure factor for a four-leg, three-band Hubbard ladder including oxygen orbitals~\cite{Huang2017} and for a $t$-$t'$-$U$ four-leg ladder~\cite{Huang2017b} have also indicated the hourglass-type excitation in the presence of the charge stripes. In contrast to hole doping, the spin-wave-like excitation persists with the introduction of electron carriers in the $t$-$t'$-$U$ Hubbard model~\cite{Huang2017b}, which is consistent with INS experiment~\cite{Fujita2012}.

Similar to the $t$-$t'$-$U$ Hubbard model, there is a clear electron-hole asymmetry in the $t$-$t'$-$J$ model which is caused by the interplay of the spin background and $t'$~\cite{Tohyama1994,Tohyama2004}. In hole doping, the ground state of a four-leg $t$-$t'$-$J$ ladder has been studied using the density-matrix renormalization group (DMRG)~\cite{Tohyama1999,White1999,Scalapino2012,Dodaroo2017}. The charge stripes are stabilized for negative $t'/t$, and by changing the sign of $t'/t$ the stripes become weaker~\cite{White1999}. The dynamical spin structure factor of the model, however, has not been studied using DMRG. 

The recent development of resonant inelastic x-ray scattering (RIXS) tuned for the Cu $L$ edge has provided a lot of new insights about spin excitations in cuprates~\cite{Ament2011,Dean2015}. Very recently, spin excitations in the so-called 1/8-doped system, La$_{1.875}$Ba$_{0.125}$CuO$_4$, were observed by RIXS, and an anomalous change in spin dispersion around $\mathbf{q}=(\pi/2,0)$ was reported below the stripe-ordered temperature~\cite{Miao2017}. The anomaly has been explained by a localized spin model reflecting the effect of the charge stripes~\cite{Miao2017}. However, there has been no investigation of spin excitations along the momentum perpendicular to the stripes based on microscopic models like the $t$-$t'$-$J$ model. Furthermore, RIXS can detect momentum-dependent charge excitations~\cite{Ishii2013,Ishii2014,Ishii2017}. Therefore, it is important to study not only spin dynamics but also charge dynamics in such a microscopic model.  

In this paper, we investigate both the dynamical spin and charge structure factors in a four-leg $t$-$t'$-$J$ ladder to give insight into momentum-dependent spin and charge dynamics in both hole-doped and electron-doped cuprates. We perform large-scale dynamical DMRG calculations. The choice of the four-leg ladder is based on the fact that (i) DMRG gives better accuracy for ladder geometry than for a purely square lattice and (ii) the spectral properties of the dynamical spin structure factor in the four-leg ladder are similar to those in a 2D system, as compared with those in a one-dimensional (1D) one as realized by comparing our results with QMC studies for the coupled Hubbard chains~\cite{Raczkowski2013,Kung2017}.

The dynamical charge structure factor along the $(0,0)$-$(\pi,0)$ direction clearly shows low-energy excitations corresponding to the stripe order in hole doping. On the other hand, the stripe order is weak in electron doping, resulting in fewer low-energy excitations, as expected. In the dynamical spin structure factor, we find incommensurate spin excitations near the magnetic zone center $\mathbf{q}=(\pi,\pi)$ forming an hourglass behavior in hole doping, while in electron doping we clearly find a spin-wave-like dispersion starting from $\mathbf{q}=(\pi,\pi)$. The hourglass behavior qualitatively agrees with the experimental data, but the high spectral weight of low-energy excitation toward $\mathbf{q}=(0,\pi)$ from the incommensurate wave vector is inconsistent with experimental observations. Along the $(0,0)$-$(\pi,0)$ direction, spin excitations are strongly influenced by the stripes in hole doping, resulting in two branches forming a discontinuous dispersion. In contrast, spin excitations in electron doping show a downward shift in energy toward $(\pi,0)$. These behaviors along the $(0,0)$-$(\pi,0)$ direction are qualitatively similar to RIXS results~\cite{Miao2017,Ishii2014}.

This paper is organized as follows. The four-leg $t$-$t'$-$J$ ladder and dynamical DMRG method are introduced  in Sec.~\ref{Sec2}. In Sec.~\ref{Sec3-0}, we calculate the charge distribution in the ground state. The dynamical charge structure factors obtained with the dynamical DMRG are compared between hole and electron dopings in Sec.~\ref{Sec3}. The dynamical spin structure factors in both hole and electron dopings are shown in Sec.~\ref{Sec4}. Finally, a summary is given in Sec.~\ref{Sec4}.

\section{Model and method}
\label{Sec2}
The Hamiltonian of the $t$-$t'$-$J$ model in two dimensions reads 
\begin{eqnarray}
H&=& -t\sum_{\mathbf{l},\boldsymbol{\delta},\sigma}
    \left( \tilde{c}_{\mathbf{l}+\boldsymbol{\delta},\sigma }^\dagger \tilde{c}_{\mathbf{l},\sigma}+\tilde{c}_{\mathbf{l}-\boldsymbol{\delta},\sigma }^\dagger \tilde{c}_{\mathbf{l},\sigma} \right) \nonumber \\
&&  -t'\sum_{\mathbf{l},\boldsymbol{\delta}',\sigma}
   \left( \tilde{c}_{\mathbf{l}+\boldsymbol{\delta}',\sigma }^\dagger \tilde{c}_{\mathbf{l},\sigma }+\tilde{c}_{\mathbf{l}-\boldsymbol{\delta}',\sigma }^\dagger \tilde{c}_{\mathbf{l},\sigma } \right) \nonumber \\
&&  +J\sum_{\mathbf{l},\boldsymbol{\delta}}
      \left( \mathbf{S}_{\mathbf{l}+\boldsymbol{\delta}}\cdot \mathbf{S}_\mathbf{l}-\frac{1}{4} n_{\mathbf{l}+\boldsymbol{\delta}} n_{\mathbf{l}}\right) ,
\label{H}
\end{eqnarray}
where $t$, $t'$, and $J$ are the nearest-neighbor hopping, the next-nearest-neighbor hopping, and the antiferromagnetic (AF) exchange interaction, respectively; $\boldsymbol{\delta}=\mathbf{x}$, $\mathbf{y}$ and $\boldsymbol{\delta}'=\mathbf{x}+\mathbf{y}$, $\mathbf{x}-\mathbf{y}$, with $\mathbf{x}$ and $\mathbf{y}$ being the unit vectors in the $x$ and $y$ directions, respectively; the operator $\tilde{c}_{\mathbf{l},\sigma}=c_{\mathbf{l},\sigma}(1-n_{\mathbf{l},-\sigma})$, with $n_{\mathbf{l},\sigma}=c_{\mathbf{l},\sigma}^\dagger c_{\mathbf{l},\sigma}$, annihilates a localized particle with spin $\sigma$ at site $\mathbf{l}$ with the constraint of no double occupancy; $\mathbf{S}_\mathbf{l}$ is the spin operator at site $\mathbf{l}$; and $n_\mathbf{l}=n_{\mathbf{l},\uparrow}+n_{\mathbf{l},\downarrow}$.  

In the model (\ref{H}), the difference between hole and electron dopings is taken into account by the sign difference of the hopping parameters~\cite{Tohyama1994}: For hole doping, the particle is an electron with $t>0$ and $t'<0$, while the particle is a hole with $t<0$ and $t'>0$ for electron doping.  We take $J/|t|=0.4$ and $t'/t=-0.25$ for both the hole- and electron-doped cases, which are typical values appropriate for cuprates with $|t|\sim0.35$~eV.

We use a $24\times 4=96$ site lattice with cylindrical geometry where the $x$ direction has an open boundary condition, while the $y$ direction has a periodic boundary condition. This lattice is called the four-leg $t$-$t'$-$J$ ladder. The carrier density for $n_\mathrm{h}$ holes ($n_\mathrm{e}$ electrons) in the ladder is defined by $x_\mathrm{h}=n_\mathrm{h}/96$ ($x_\mathrm{e}=n_\mathrm{e}/96$).  In the $24\times 4$ ladder ($L_x=24$ and $L_y=4$), the $y$ component of momentum $\mathbf{q}$ is determined by using standard translational symmetry, i.e., $q_y=2n_y\pi/L_y$ ($n_y=0, \pm 1, L_y/2$), but the $x$ component is given by $q_x=n_x\pi/(L_x+1)$ ($n_x=1,2,\cdots,L_x$) because of the open boundary condition. Defining $l_x$ ($l_y$) as the $x$ ($y$) component of site $\mathbf{l}$, we can write the Fourier component of the charge operator and that of the $z$ component of the spin operator as 
\begin{equation}
N_\mathbf{q}=\sqrt{\frac{2}{(L_x+1)L_y}} \sum_\mathbf{l} \sin(q_x l_x) e^{-iq_y l_y}n_\mathbf{l}
\label{N}
\end{equation}
and
\begin{equation}
S_\mathbf{q}^z=\sqrt{\frac{2}{(L_x+1)L_y}} \sum_\mathbf{l} \sin(q_x l_x) e^{-iq_y l_y}S_\mathbf{l}^z \;,
\label{Sz}
\end{equation}
respectively. 

The dynamical charge and spin structure factors, $N(\mathbf{q},\omega)$ and $S(\mathbf{q},\omega)$, are defined as
\begin{eqnarray}
N(\mathbf{q},\omega)&=&-\frac{1}{\pi} \mathrm{Im} \left\langle 0 \right| \tilde{N}_{-\mathbf{q}} \frac{1}{\omega  - H + E_0+i\gamma } \tilde{N}_\mathbf{q} \left| 0 \right\rangle \label{Nqw}\\
S(\mathbf{q},\omega)&=&-\frac{1}{\pi} \mathrm{Im} \left\langle 0 \right| S_{-\mathbf{q}}^z \frac{1}{\omega  - H + E_0+i\gamma } S_\mathbf{q}^z \left| 0 \right\rangle \label{Sqw},
\end{eqnarray}
where $\left|0 \right\rangle$ represents the ground state with energy $E_0$,  $\tilde{N}_\mathbf{q}=N_\mathbf{q}-\left\langle 0\right| N_\mathbf{q} \left| 0\right\rangle$, and $\gamma$ is a small positive number.

We calculate Eqs.~(\ref{Nqw}) and (\ref{Sqw}) for the $24\times 4$ $t$-$t'$-$J$ ladder using dynamical DMRG, where we use three kinds of target states: for $N(\mathbf{q},\omega)$, (i) $\left| 0\right\rangle$, (ii) $\tilde{N}_\mathbf{q} \left| 0 \right\rangle$, and (iii) $(\omega - H + E_0+i\gamma)^{-1} \tilde{N}_\mathbf{q} \left| 0 \right\rangle$. Target state (iii) is evaluated using a kernel-polynomial expansion method~\cite{Sota2010}, where the Lorentzian broadening $\gamma$ in Eqs.~(\ref{Nqw}) and  (\ref{Sqw}) is replaced by a Gaussian broadening with a half width at half maximum of $0.08|t|$. In our numerical calculations, we divide the energy interval $[0,2|t|]$ by 100 mesh points and target all of the points at once. To perform DMRG, we construct a snakelike one-dimensional chain and use the maximum truncation number $m=4000$, and the resulting truncation error is less than $3\times10^{-4}$. To check the effect of the leg length $L_x$ on $S(\mathbf{q},\omega)$, we performed dynamical DMRG calculations with $m=2000$ for a $12\times 4$ site ladder. We found that the results for the $L_x=12$ system lead to the same conclusions as the case for the $L_x=24$ system, except for the sparseness of $q_x$ defined in the BZ.

\section{Charge distribution}
\label{Sec3-0}

\begin{figure}[tb]
\epsfig{file=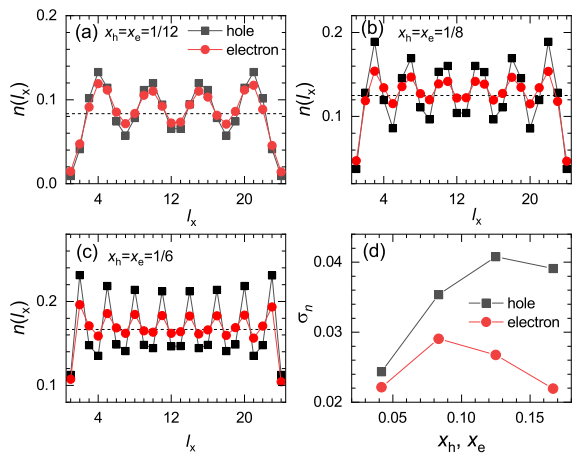, width=75mm}
\caption{Carrier distribution in the $24\times 4$ $t$-$t'$-$J$ ladder. $t=1(-1)$, $t'=-0.25 (0.25)$, and $J=0.4$ for hole (electron) doping. The carrier number $n(l_x)$ along the leg position $l_x$ for (a) $x_\mathrm{h}=x_\mathrm{e}=1/12$, (b) $x_\mathrm{h}=x_\mathrm{e}=1/8$, and (c) $x_\mathrm{h}=x_\mathrm{e}=1/6$. The black squares (red circles) represent $n(l_x)$ for hole (electron) doping. The horizontal dotted line denotes the averaged number $x_\mathrm{h}=x_\mathrm{e}$. (d) The $x_\mathrm{h}$ ($x_\mathrm{e}$) dependence of the standard deviation $\sigma_n$ for $n(l_x)$ in hole (electron) doping.}
\label{fig1}
\end{figure}

We first examine the carrier distribution in the ground state of the four-leg $t$-$t'$-$J$ ladder to confirm the nature of the charge stripes reported previously by DMRG~\cite{Tohyama1999,White1999,Scalapino2012,Dodaroo2017}. Figure~\ref{fig1} shows the carrier number $n(l_x)$ along the leg position $l_x$. Note that there is no carrier number dependence on the rung position $l_y$. As expected, there is an oscillation of $n(l_x)$ in the middle of the ladder, depending on the charge density in both hole and electron dopings. The period is six-, four-, and three-lattice spacing for $x_\mathrm{h}=x_\mathrm{e}=1/12$, 1/8, and 1/6, as shown in Figs.~\ref{fig1}(a), \ref{fig1}(b), and \ref{fig1}(c), respectively, implying that the period is given by $(2x_{\mathrm{h}(\mathrm{e})})^{-1}$. The amplitude of the oscillation is smaller in electron doping than in hole doping. This has been pointed out in the context of a sign change of $t'$, where positive $t'$ suppresses the stripes~\cite{White1999}. The standard deviation $\sigma_n$ for $n(l_x)$ is plotted as a function of the carrier number in Fig.~\ref{fig1}(d). In hole doping, $\sigma_n$ has a maximum at $x_\mathrm{h}=1/8$, implying the strongest stripe order near the $1/8$ doping, as observed in hole-doped cuprates~\cite{Tranquada1995}. Such a stripe order is organized by the $J$ term in (\ref{H}). In fact, we found that $\sigma_n$ at $x_\mathrm{h}=1/8$ becomes almost equal to that at $x_\mathrm{h}=1/6$ for $J=0.2$, and with further reducing $J$, $\sigma_n$ at $x_\mathrm{h}=1/8$ decreases and becomes almost zero at $J=0$, i.e., no charge inhomogeneity (not shown). In electron doping, $\sigma_n$ is small and decreases above $x_\mathrm{e}=1/12$. 

In contrast to the nonuniform carrier density, an expectation value of spin density on each site is zero in our DMRG calculations since the calculations preserve the rotational symmetry of spin space. If one introduces an external magnetic field at the edges, the symmetry is broken, and thus, local spin density becomes finite, as discussed for previous DMRG calculations~\cite{White1999}. In hole doping, there is a ferromagnetic spin arrangement across the charge stripes, resulting in antiphase spin structures~\cite{White1999,Scalapino2012}.

\section{Dynamical charge structure factor}
\label{Sec3}

\begin{figure}[tb]
\epsfig{file=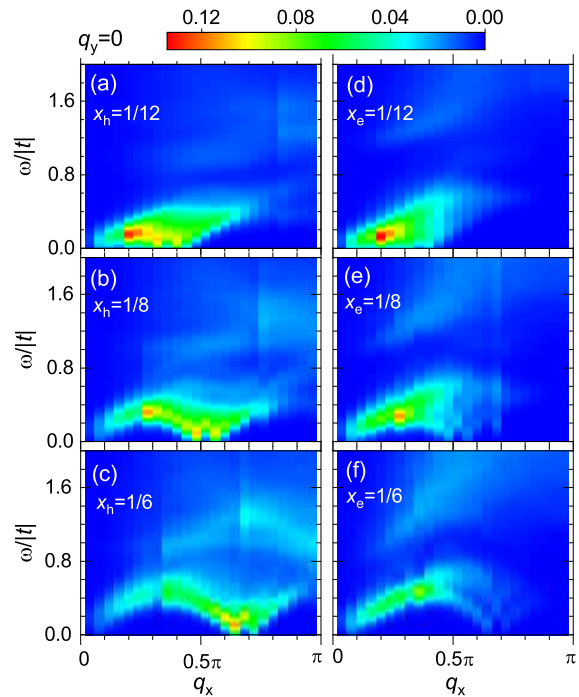, width=75mm}
\caption{$N(\mathbf{q},\omega)$ along the $(0,0)$-$(\pi,0)$ direction in the $24\times 4$ $t$-$t'$-$J$ ladder with $J=0.4$. (a) $x_\mathrm{h}=1/12$, (b) $x_\mathrm{h}=1/8$, and (c) $x_\mathrm{h}=1/6$ for hole doping ($t=1$ and $t'=-0.25$). (d) $x_\mathrm{e}=1/12$, (e) $x_\mathrm{e}=1/8$, and (f) $x_\mathrm{e}=1/6$ for electron doping ($t=-1$ and $t'=0.25$).}
\label{fig2}
\end{figure}

Figure~\ref{fig2} shows $N(\mathbf{q},\omega)$ along the $(0,0)$-$(\pi,0)$ direction for both hole and electron dopings. As expected from the stripe ground state, strong low-energy excitations whose energy minimum is located around $q_x=4x_\mathrm{h}\pi$ emerge in the hole-doped case [see Figs.~\ref{fig2}(a), \ref{fig2}(b), and \ref{fig2}(c)]. There are broad but weak excitations at $\omega<0.8=2J$. In contrast to the hole doping, low-energy excitations near $q_x=4x_\mathrm{e}\pi$ for electron doping [see Figs.~\ref{fig2}(d), \ref{fig2}(e), and \ref{fig2}(f)] are very weak, reflecting weak stripe ordering as discussed above. High-energy excitations above $\omega=0.8|t|$ show broad dispersive features that are steeper than those in hole doping. Such dispersive high-energy excitations have been observed in the Cu $L$-edge RIXS for electron-doped cuprates~\cite{Ishii2014,Lee2014}. 

\begin{figure*}[tb]
\epsfig{file=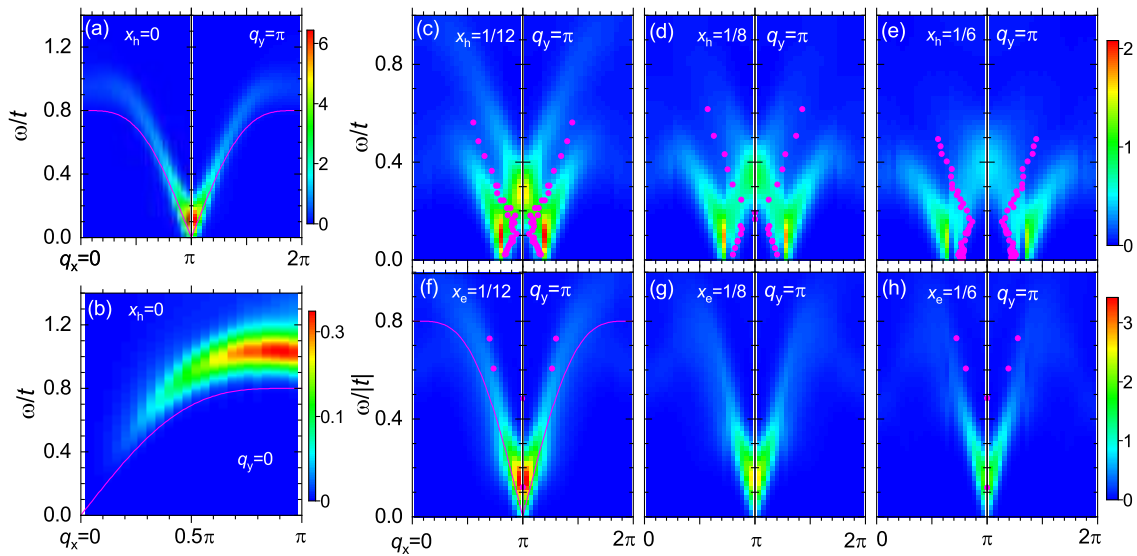, width=150mm}
\caption{$S(\mathbf{q},\omega)$ in the $24\times 4$ $t$-$t'$-$J$ ladder with $J=0.4$. (a) Half filling ($x_\mathrm{h}=0$) and $q_y=\pi$. (b) Half filling and $q_y=0$. (c), (d), and (e) Hole doping ($t=1$ and $t'=-0.25$) with $x_\mathrm{h}=1/12$, $1/8$, and $1/6$, respectively, along the $(0,\pi)$-$(2\pi,\pi)$ direction. The purple dots in (c), (d), and (e) represent the peak position of INS experiments for La$_{1.915}$Sr$_{0.085}$CuO$_4$~\cite{Lipscombe2009}, La$_{1.875}$Ba$_{0.125}$CuO$_4$~\cite{Tranquada2004}, and La$_{1.84}$Sr$_{0.16}$CuO$_4$~\cite{Vignolle2007}, respectively, where we assume $J=132$~meV~\cite{Hayden1991}. (f), (g), and (h) Electron doping ($t=-1$ and $t'=0.25$) with $x_\mathrm{e}=1/12$, $1/8$, and $1/6$, respectively, along the $(0,\pi)$-$(2\pi,\pi)$ direction. The purple dots in (f) and (h) represent the peak position of INS experiments for Pr$_{1.4-x}$La$_{0.6}$Ce$_x$CuO$_{4+\delta}$~\cite{Asano2017} with $x=0.08$ and $x=0.18$, respectively. The purple lines in (a), (b), and (f) represent a single magnon dispersion at half filling obtained with the linear spin-wave theory for the 2D Heisenberg model.}
\label{fig3}
\end{figure*}

Strong low-energy excitations for small $\mathbf{q}$ in the low-doping region in $N(\mathbf{q},\omega)$ have been proposed on the electron-doped side of the 2D Hubbard model with $t'$~\cite{Tohyama2015}. The origin of the strong intensity has been attributed to the proximity to the phase separation from the study of the $t$-$t'$-$J$ model~\cite{Greco2017}. Such strong intensity is also seen at $x_\mathrm{e}=1/12$ in Fig.~\ref{fig2}(d). The intensity is, in fact, reduced by introducing the nearest-neighbor Coulomb interaction (not shown here) as demonstrated in \cite{Greco2017}.    

\section{Dynamical spin structure factor}
\label{Sec4}

\subsubsection{The $(0,\pi)-(\pi,\pi)$ direction}

At half filling ($x_\mathrm{h}=0$), the four-leg $t$-$t'$-$J$ ladder exhibits a spin gap whose magnitude is close to $0.2J$~\cite{White1994}. In Fig.~\ref{fig3}(a), the gap is identified as the peak position of $S(\mathbf{q},\omega)$ at $\mathbf{q}=(\pi,\pi)$, which is close to $0.2J=0.08t$. A spin-wave-like dispersion exists toward $\mathbf{q}=(0,\pi)$ from $(\pi,\pi)$ and along the $(0,0)$-$(\pi,0)$ direction [Fig.~\ref{fig3}(b)], whose energy is slightly higher than the energy of dispersion obtained by the linear-spin-wave theory for a 2D Heisenberg model [purple lines in Figs.~\ref{fig3}(a) and \ref{fig3}(b)].

With hole doping, the $\mathbf{q}=(\pi,\pi)$ excitation at half filling splits into two low-energy excitations along the $(0,\pi)$-$(\pi,\pi)$ direction, as shown in Figs.~\ref{fig3}(c), \ref{fig3}(d), and \ref{fig3}(e). The wave vector measured from $\mathbf{q}=(\pi,\pi)$ is approximately given by $(\pm 2x_\mathrm{h}\pi,0)$, which is consistent with incommensurate vectors reported in hole-doped cuprate superconductors La$_{2-x}$Sr$_x$CuO$_4$~\cite{Yamada1998}. Linear dispersive branches emerge from the $q_x$ position toward both the $q_x=\pi$ (inward) and $q_x=0$ (outward) directions in all three densities, $x_\mathrm{h}=1/12$, 1/8, and 1/6. 

In the INS experiment~\cite{Fujita2012}, the outward dispersion has not been observed. Furthermore, in other calculations of $S(\mathbf{q},\omega)$ under the stripe order for the 2D extended Hubbard model based on the random-phase approximation~\cite{Kaneshita2001} and time-dependent Gutzwiller approximation (TDGA)~\cite{Seibold2006}, the outward dispersion loses its intensity quickly for small $x_\mathrm{h}$. However, the outward dispersion is clearly seen in the $t$-$t'$-$J$ ladder. This inconsistency may arise from ladder geometry in our model, which is different from 2D geometry in the experiment and other calculations. To confirm this, we need to perform the calculation of $S(\mathbf{q},\omega)$ for a square $t$-$t'$-$J$ lattice. This remains to be a future problem.

The inward dispersive structure merges with that from the opposite side and forms an intense structure at $\mathbf{q}=(\pi,\pi)$, whose energy position is clearly lower than $J$, i.e., $\omega\sim 0.7J=0.28t$ at $x_\mathrm{h}=1/12$ and increases with increasing $x_\mathrm{h}$ up to $\omega\sim J=0.4t$ at $x_\mathrm{h}=1/6$. Above the $(\pi,\pi)$ structure there is no gap, in contrast to the TDGA results for the 2D extended Hubbard model~\cite{Seibold2006}. Rather, there is a linear dispersive structure with low intensity extending, for example, up to $\omega\sim t$ at $x_\mathrm{h}=1/12$, as seen in Fig.~\ref{fig3}(c). The linear dispersive feature looks to be continuously connected to the linear dispersions starting from the incommensurate position. Such behavior has been reported in a linear spin-wave theory for a spin model assuming a bond-centered vertical stripe, where AF exchange interaction is assumed for every nearest-neighbor bond except for the ferromagnetic bonds across the stripe~\cite{Carlson2004}. In this view, the upward-energy shift of the strong-intensity position at $(\pi,\pi)$ with increasing $x_\mathrm{h}$ can be partly related to the outward shift of the incommensurate wave vectors, where we assume that the velocity of spin-wave-like dispersion starting from the incommensurate points does not change significantly with $x_\mathrm{h}$.

In Figs.~\ref{fig3}(c), \ref{fig3}(d), and \ref{fig3}(e),  the calculated spectra are compared with the experimental peak positions~\cite{Lipscombe2009,Tranquada2004,Vignolle2007}, assuming $J=132$~meV~\cite{Hayden1991}. The neck position of the hourglass dispersions is lower in energy than the position of the calculated $(\pi,\pi)$ spectrum. The difference can be attributed to the ladder geometry whose spectral weight tends to shift to a higher-energy position compared with 2D systems, as demonstrated by the comparison with the linear spin-wave theory at half filling [see Figs.~\ref{fig3}(a) and \ref{fig3}(b)].

In contrast to hole doping, the lowest-energy excitation of $S(\mathbf{q},\omega)$ in electron doping remains at $\mathbf{q}=(\pi,\pi)$ for all $x_\mathrm{e}$, as shown in Figs.~\ref{fig3}(f), \ref{fig3}(g), and \ref{fig3}(h). Such low-energy excitations at $(\pi,\pi)$ are expected from strong AF correlation in electron doping due to the effect of $t'$~\cite{Tohyama1994,Tohyama2004}. The dispersive behavior near $(\pi,\pi)$ is similar to that at half filling, but away from $(\pi,\pi)$ the spectral distribution becomes broader compared with that in Fig.~\ref{fig3}(a). This is qualitatively consistent with INS experiments for electron-doped cuprates~\cite{Fujita2012,Asano2017,Wilson2006,Fujita2006}. The experimental peak positions for Pr$_{1.4-x}$La$_{0.6}$Ce$_x$CuO$_{4+\delta}$~\cite{Asano2017} are plotted in Figs.~\ref{fig3}(f) and \ref{fig3}(h). We find a rough agreement with our calculated results. 

\begin{figure}[tb]
\epsfig{file=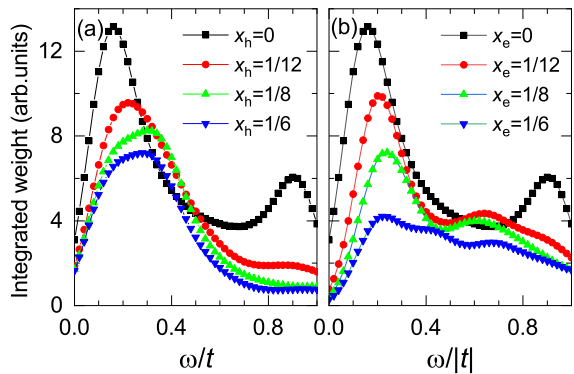, width=75mm}
\caption{Integrated weight of $S(\mathbf{q},\omega)$ with respect to $\mathbf{q}$ along the $(0.\pi)$-$(\pi,\pi)$ direction for the $24\times 4$ $t$-$t'$-$J$ ladder with $J/|t|=0.4$ and $t'/t=-0.25$. (a) Hole doping and (b) electron doping.}
\label{fig4}
\end{figure}

Since spectral weight in the INS experiments is concentrated on the region around $\mathbf{q}=(\pi,\pi)$, integrated weight around $\mathbf{q}=(\pi,\pi)$ has been analyzed in the experimental literature~\cite{Tranquada2004,Fujita2006,Wakimoto2007}. To make a possible comparison with the experimental data, we show in Fig.~\ref{fig4} the integrated weight of $S(\mathbf{q},\omega)$ with respect to $\mathbf{q}$ along the $(0,\pi)$-$(\pi,\pi)$ direction for both hole and electron dopings. In hole doping, the lowest-energy peak at $\omega=0.15t$ decreases in weight with $x_\mathrm{h}$ and broadens with increasing the weight at higher energy. At $x_\mathrm{h}=1/8$, a new peak appears at $\omega\sim 0.3t$, which is consistent with the experiments~\cite{Tranquada2004,Wakimoto2007}. The energy region higher than $\omega\sim 0.6t$ loses weight significantly. In contrast, the weight in the high-energy region in electron doping remains less $x_\mathrm{e}$ dependent, as shown in Fig.~\ref{fig4}(b). We also notice that a high-energy peak at $\omega\sim 0.9|t|$ at half filling $x_\mathrm{e}=0$ shifts to the lower-energy side around $\omega\sim 0.7|t|$ with increasing $x_\mathrm{e}$. This corresponds to the broadening of the spin-wave-like dispersion, as discussed above. Since the weight around $\omega\sim 0.2t$ decreases quickly with $x_\mathrm{e}$, it is smaller than that for hole doping above $x_\mathrm{e}=1/8$, being qualitatively similar to the experiment~\cite{Fujita2006}.

\begin{figure}[tb]
\epsfig{file=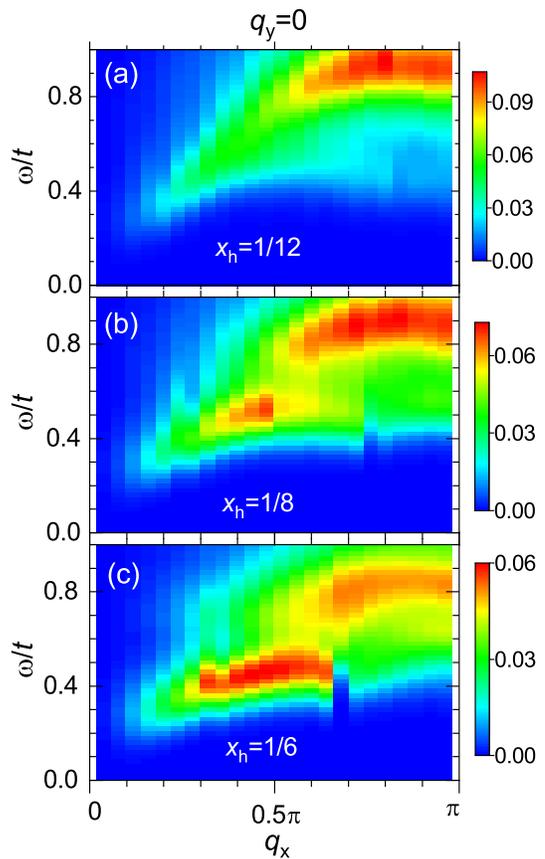, width=70mm}
\caption{$S(\mathbf{q},\omega)$ along the $(0,0)$-$(\pi,0)$ direction in the hole-doped $24\times 4$ $t$-$t'$-$J$ ladder with $J=0.4$. (a) $x_\mathrm{h}=1/12$, (b) $x_\mathrm{h}=1/8$, and (c) $x_\mathrm{h}=1/6$.}
\label{fig5}
\end{figure}

\subsubsection{The $(0,0)-(\pi,0)$ direction}

Since the charge stripe has a charge modulation along the $(0,0)$-$(\pi,0)$ direction, $S(\mathbf{q},\omega)$ is expected to show spectral features associated with the stripe. In fact, $S(\mathbf{q},\omega)$ for hole doping clearly exhibits such a feature, as shown in Fig.~\ref{fig5}. For $x_\mathrm{h}=1/8$ (1/6), discontinuous spectral intensity appears at $q_x\sim 4x_\mathrm{h}\pi$ close to the stripe wave vector, as seen in Fig.~\ref{fig5}(b) [Fig.~\ref{fig5}(c)]. More precisely, there are two branches, one of which has low-energy excitations with maximum energy $\omega\sim 0.5t$ near $q_x=0.5\pi$ and the other of which exhibits high-energy excitations around $\omega\sim 0.8t$. 

The anomaly at $q_x\sim 4x_\mathrm{h}$ in spin excitation has been reported in RIXS for La$_{1.875}$Ba$_{0.125}$CuO$_4$~\cite{Miao2017}, although clear discontinuous spectral intensity has not been identified. In the interpretation of the experimental data, a localized spin model has been introduced, where AF magnetic exchange interaction across disordered charge stripes is replaced by a ferromagnetic one~\cite{Miao2017}. A simple view of the presence of two branches in our results is also given by a one-dimensional spin model where a ferromagnetic exchange interaction is periodically introduced onto one of two bonds. In this simple model, the two branches show an anticrossing, leading to a gap at the middle of the magnetic BZ and a clear separation of the two branches. Therefore, the two separated branches obtained with our dynamical DMRG calculations indicate the presence of ferromagnetic effective interaction along the perpendicular direction of the charge stripes. The presence of effective ferromagnetic interaction is evidenced by ferromagnetically aligned spins in the hole-rich region~\cite{White1999,Scalapino2012}.

\begin{figure}[tb]
\epsfig{file=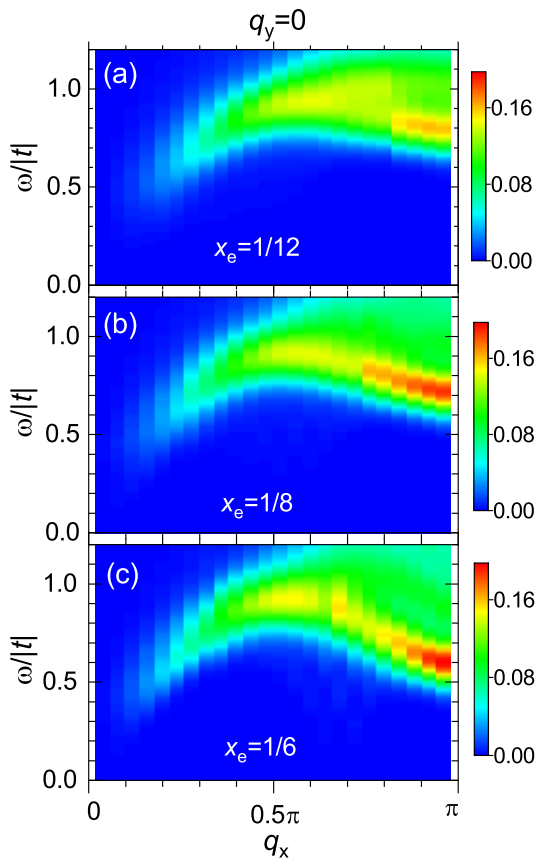, width=70mm}
\caption{$S(\mathbf{q},\omega)$ along the $(0,0)$-$(\pi,0)$ direction in the electron-doped $24\times 4$ $t$-$t'$-$J$ ladder with $J=0.4$. (a) $x_\mathrm{e}=1/12$, (b) $x_\mathrm{e}=1/8$, and (c) $x_\mathrm{e}=1/6$.}
\label{fig6}
\end{figure}

In electron doping, such a discontinuous behavior of spectral weights is invisible in $S(\mathbf{q},\omega)$, as shown in Fig.~\ref{fig6}. This is consistent with weak charge stripe moderation in electron doping, as discussed in Sec.~\ref{Sec3}. Alternatively, one can find a peculiar spectral behavior in contrast to hole doping, which is a downward shift of the peak position of spectral weight beyond $q_x\sim 0.5\pi$ for all three $x_\mathrm{e}$ cases. This is a counterintuitive behavior in the sense that spin excitations similar to the Heisenberg model might be expected as evidenced from the spin-wave-like dispersions near $\mathbf{q}=(\pi,\pi)$. This downward behavior is thus due to the presence of electron carriers, suggesting the influence of the itinerant nature in the electron-doped system. The signature of such a downward shift has not clearly  been seen in the experimental data of RIXS for Nd$_{2-x}$Ce$_x$CuO$_4$~\cite{Ishii2014,Lee2014}. However, it is clear that the dispersion along the $(0,0)$-$(\pi,0)$ direction in electron-doped cuprates~\cite{Ishii2014} becomes flat above $(\pi/2,0)$, in contrast to hole-doped cuprates with monotonically increasing dispersion~\cite{LeTacon2011,LeTacon2013,Dean2013}.

The spin-excitation energy at $\mathbf{q}=(\pi/2,0)$ in Fig.~\ref{fig6} remains almost the same as that at half filling in Fig.~\ref{fig3}(b). This is different from the experimental data, where the energy increases with increasing $x_\mathrm{e}$~\cite{Ishii2014,Lee2014}. This difference will disappear if one introduces the so-called three-site terms into the $t$-$t'$-$J$ model~\cite{Jia2014}; that is, the terms shift spectral weight at higher energy. In order to clarify the effect of the three-site terms in ladder geometry, we performed a Lanczos-type exact diagonalization (ED) calculation of $S(\mathbf{q},\omega)$ and $N(\mathbf{q},\omega)$ for a $5\times 4$ cylindrical $t$-$t'$-$J$ ladder with the three-site terms (not shown). We found that the spectral weights shift to higher energy almost independent of $\mathbf{q}$ but their spectral shapes are qualitatively unchanged. This suggests that the conclusions in this paper do not change in the presence of the three-site terms for the $24\times 4$ ladder. We note that the dynamical DMRG  calculation with the three-site terms remains to be a future problem.

Finally, we discuss the effect of dimensionality on $S(\mathbf{q},\omega)$. The question remains whether the results of the four-leg ladder are close to those of the square lattice or two-leg ladder. Comparing our results with $S(\mathbf{q},\omega)$ of two-leg ladders, the latter of which has been reported for a $10\times 2$ periodic $t$-$J$ ladder by ED~\cite{Troyer1996}, a $16\times 2$ periodic $t$-$J$ ladder by reduced Hilbert space ED~\cite{Dagotto1998}, a $48\times 2$ cylindrical $t$-$U$-$J$ ladder by dynamical DMRG~\cite{Nocera2017}, and a $24\times 2$ cylindrical $t$-$t'$-$J$ ladder by our dynamical DMRG (not shown), we can find the differences  between the two-leg and four-leg ladders. One of the significant differences is the incommensurate wave vector $\tilde{q}_x$ away from $\mathbf{q}=(\pi,\pi)$, where $\tilde{q}_x\sim \pm 2x_\mathrm{h}\pi$ for the four-leg ladder, while $\tilde{q}_x\sim \pm x_\mathrm{h}\pi$ for the two-leg ladder~\cite{Nocera2017}. The $\tilde{q}_x$ for the four-leg ladder is a consequence of the stripes, which is absent in the two-leg ladder but appears in the square lattice of the $t$-$t'$-$U$ Hubbard model, as demonstrated, for example, by the variational Monte Carlo calculation~\cite{Ido2018}. Another difference is seen on $S(\mathbf{q},\omega)$ along  $\mathbf{q}=(q_x,0)$: in the four-leg ladder the spectral weights near $q_x=\pi$ are negligible below $\omega\sim J$, while in the two-leg ladder significant weights exist in the energy region~\cite{Troyer1996,Nocera2017}. Such a difference is also seen in the calculations of the 1D-2D crossover for the coupled Hubbard chains; that is, the low-energy excitation near $\mathbf{q}=(\pi,0)$ has strong intensity in the quasi-1D case~\cite{Raczkowski2013,Kung2017}.

\section{Summary}
\label{Sec5}
In summary, we have investigated the dynamical spin and charge structure factors, $S(\mathbf{q},\omega)$ and $N(\mathbf{q},\omega)$, in the four-leg $t$-$t'$-$J$ ladder using dynamical DMRG. $N(\mathbf{q},\omega)$ along the $(0,0)$-$(\pi,0)$ direction clearly shows the low-energy excitations corresponding to the stripe order in hole doping, while the stripe order weakens in electron doping, resulting in fewer low-energy excitations. In $S(\mathbf{q},\omega)$, we found incommensurate spin excitations near the magnetic zone center $\mathbf{q}=(\pi,\pi)$ displaying an hourglass behavior in hole doping. However, the outward dispersion from the incommensurate position is strong in intensity, inconsistent with INS experiments. In electron doping, clear spin-wave-like dispersions starting from  $\mathbf{q}=(\pi,\pi)$ were seen and were similar to INS experiments. Along the $(0,0)$-$(\pi,0)$ direction, the spin excitations are strongly influenced by the stripes in hole doping, resulting in two branches that form a jump in the dispersion. In contrast, the spin excitations show a downward shift in energy toward $(\pi,0)$. These behaviors along the $(0,0)$-$(\pi,0)$ direction are also qualitatively consistent with RIXS results. For more quantitative descriptions, we need to treat 2D systems rather than ladder systems. This remains to be a future problem.

\begin{acknowledgments}
We thank M. Fujita for providing us experimental data and for fruitful discussions. We also thank K. Ishii for useful discussions. This work was supported by MEXT, Japan, as a social and scientific priority issue (creation of new functional devices and high-performance materials to support next-generation industries) to be tackled by using a post-K computer, by MEXT HPCI Strategic Programs for Innovative Research (SPIRE; hp170114), and by the interuniversity cooperative research program of IMR, Tohoku University. The numerical calculation was carried out at the K Computer and HOKUSAI, RIKEN Advanced Institute for Computational Science, and the facilities of the Supercomputer Center, Institute for Solid State Physics, University of Tokyo. This work was also supported by the Japan Society for the Promotion of Science, KAKENHI (Grants No. 26287079 and No. 15K05192).
\end{acknowledgments}



\end{document}